\documentclass[reprint,aps,prl,amsmath,amssymb,superscriptaddress,nofootinbib]{revtex4-2}

\usepackage{graphicx}
\usepackage{xcolor}
\usepackage[hidelinks]{hyperref}

\newcommand{\orcid}[1]{\begingroup
  \hypersetup{hidelinks}\href{https://orcid.org/#1}{\includegraphics[width=10pt]{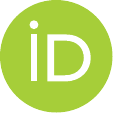}} \endgroup}

\makeatletter
\AtBeginDocument{\@booleantrue\acknowledgments@sw}
\makeatother

\begin{document}
\title{Comment on ``\textit{Fermi}-LAT Galactic Center Excess Morphology of
Dark Matter in Simulations of the Milky Way Galaxy''}

\author{Kevork N.\ Abazajian\orcid{0000-0001-9919-6362}}
\email{kevork@uci.edu}
\affiliation{Department of Physics and Astronomy, University of California --
Irvine, Irvine, California 92697-4575, USA}

\author{Jason Kumar\orcid{0009-0003-1230-2276}}
\email{jkumar@hawaii.edu}
\affiliation{Department of Physics and Astronomy, University of Hawai`i,
Honolulu, Hawaii 96822, USA}

\author{Oscar Macias\orcid{0000-0001-8867-2693}}
\email{macias@sfsu.edu}
\affiliation{Department of Physics and Astronomy, San Francisco State University, San Francisco, CA 94132, USA}
\date{\today}

\begin{abstract}
Muru \textit{et al.} [Phys.\ Rev.\ Lett.\ \textbf{135}, 161005 (2025)] conclude from the HESTIA constrained simulations that the $\gamma$-ray morphologies of the Galactic Center Excess predicted by dark matter annihilation and by an old stellar population are essentially indistinguishable.
Their ``quadratic dark matter'' maps, however, show the square of the mass projected along the line of sight, $\bigl[\int\rho\,s^{2}\,ds\bigr]^{2}$, rather than the annihilation integral $\int\rho^{2}\,ds$.
Squaring a nonnegative map preserves its isophote shapes at corresponding intensity levels, as confirmed by the reported contour axis ratios.
The corrected calculation on the same simulations does not support their conclusion.
\end{abstract}

\maketitle

Muru \textit{et al.}~\cite{Muru:2025vpz} compare the projected dark matter (DM) and old stellar distributions in six Milky Way analogs from the HESTIA constrained simulations~\cite{2020MNRAS.498.2968L}. They conclude that DM annihilation and an old stellar population predict essentially indistinguishable $\gamma$-ray morphologies for the Galactic Center Excess. However, their ``quadratic DM'' maps are obtained by squaring the projected mass, rather than by integrating the squared density along the line of sight.

The velocity-independent annihilation intensity is proportional to the line-of-sight integral $J(b,\ell)=\int\rho^{2}(s,b,\ell)\,ds$. The procedure in Appendix~A of Ref.~\cite{Muru:2025vpz} instead sums particle masses within a $3^{\circ}$ aperture around each line of sight, retaining particles out to $15$~kpc from the observer. It then squares this map, proportional to $\int\rho\,s^{2}ds$, as implemented in the public code.\footnote{\url{https://gitlab.aip.de/muru/gce_in_hestia}} The resulting quantity is $\bigl[\int\rho\,s^{2}\,ds\bigr]^{2}$, not $\int\rho^{2}(s)\,ds$.

Annihilation requires squaring the density before integration. The projected mass also contains an $s^{2}$ volume factor that cancels against the $1/s^{2}$ flux dilution in an intensity calculation. For emissivity $\epsilon$, this leaves $\int\epsilon\,ds$. Annihilation has $\epsilon\propto\rho^{2}$, whereas the assumed emission from old stars has $\epsilon\propto\rho_{\star}$. The stellar maps therefore also require the inverse-square distance weighting.

Squaring a nonnegative map preserves its isophote shapes. The contour of $f^{2}$ at level $c^{2}$ encloses the same region as the contour of $f$ at level $c$, so their axis ratios are identical. This accounts for the matching axis ratios of the DM and quadratic-DM panels in Figs.~1--2 of Ref.~\cite{Muru:2025vpz} at level pairs $(q,q^{2})$ in every galaxy. Although squaring changes relative intensities, these matching contours are not a separate prediction for annihilation emission.

Using the same particles, we calculate $J\Delta\Omega=\sum_{i}m_{i}\rho_{i}/s_{i}^{2}$ per beam, with local densities from a $k$-nearest-neighbor estimate. The stellar maps omit the factor $\rho_{i}$. For 36 observer azimuths per halo (216 views in total), we measure the concentration $C\equiv F(<3^{\circ})/F(<10^{\circ})$ in circular apertures under the flat-sky approximation, after applying Gaussian smoothing with full width at half maximum $2^{\circ}$. The median concentration is $0.43$ for annihilation and $0.23$ for the simulated stellar maps. The Freudenreich~\cite{Freudenreich:1997bx}, Cao~\cite{Cao:2013dwa}, and Coleman~\cite{Coleman:2019kax} bulge templates give $0.27$, $0.27$, and $0.29$, respectively. Every annihilation map is more concentrated than each template and its corresponding stellar map.

Our corrected annihilation maps remain flattened in the same direction as the bulge, but we measure a median minor-to-major axis ratio of $0.76$, larger than the $0.72$, $0.56$, and $0.62$ of the three bulge templates. We calculate these ratios from the positions of pixels
with intensities above $30\%$ of each map's maximum, without
intensity weighting. The larger axis ratio and greater central concentration are consistent with the $\rho^{2}$ weighting, which emphasizes the inner halo where Ref.~\cite{Muru:2025vpz} finds less flattening.
Observational analyses have also examined departures from spherical symmetry in the excess, including tests of an oblate profile with $c/a=0.7$ in Ref.~\cite{Abazajian:2020tww}. Abazajian and
Kaplinghat~\cite{Abazajian:2012pn} found that the early data could not distinguish spherical models from axisymmetric models with axis ratio $1{:}2$ (more flattened than any morphology in Ref.~\cite{Muru:2025vpz}). With a larger data set, Daylan \textit{et
al.}~\cite{Daylan:2014rsa} found the excess close to spherical.

In this Comment, we compare the predicted morphologies of annihilation and stellar emission after correcting the line-of-sight projection. Even after Gaussian smoothing to $3^{\circ}$, every annihilation map remains more concentrated than its stellar counterpart. For G3.1, our linear DM projection agrees with the public code's output to relative error $<10^{-6}$ when both use the same test geometry. We test the corrected estimator separately on particles drawn from a Navarro--Frenk--White profile, comparing the result with the analytic $\int\rho^{2}ds$.
The Supplementary Material describes these checks and the construction of the maps in Fig.~\ref{fig:maps}.

The simulations do not resolve density structure below their $0.22$~kpc gravitational softening length ($1.6^{\circ}$ at the solar circle), and smoothing does not remove the contribution of unresolved emission to the central flux. 
Regardless of resolution, annihilation requires integrating the squared density rather than squaring a projected mass. With the correct line-of-sight weighting, the same halos yield annihilation maps that remain flattened but are more centrally concentrated than their stellar counterparts. Halo flattening alone therefore does not establish the morphological equivalence claimed in Ref.~\cite{Muru:2025vpz}.

\begin{acknowledgments}
We thank N.~I.~Libeskind and M.~M.~Muru for providing the particle data for the six halos.
KNA is partially supported by the U.S. National Science Foundation (NSF) Theoretical Physics Program Grant No.\ PHY-2609954.
JK is supported in part by DOE grant DE-SC0010504. 
OM is supported by the U.S. National Science Foundation under Grant No. 2418730. 
The authors used Claude Code (Anthropic), running the Claude Opus~5, Fable~5, and Fable~5.1 models, to implement and execute the numerical analysis and to draft portions of the text.
ChatGPT (OpenAI) assisted with checks of the supplied code and derived results and with manuscript revision.
The authors specified the analysis design and numerical validation tests.
\end{acknowledgments}

\section*{Data availability}
The HESTIA particle data and the AHF halo-finder profiles analyzed here
were provided by the authors of Ref.~\cite{Muru:2025vpz} and are available
from them on reasonable request, as stated in that work; we do not
redistribute them. The analysis code, the validation against the public
code of Ref.~\cite{Muru:2025vpz}
(\url{https://gitlab.aip.de/muru/gce_in_hestia}), and the derived products
from which every number in this Comment can be recomputed without the
particle data---the projected maps, the morphology statistics for all 216
viewing geometries, the comparison templates, and the viewing frames---are
archived at Zenodo (DOI: \texttt{10.5281/zenodo.22836567}) and maintained at
\url{https://github.com/kabazajian/gce-hestia-comment}. The bulge templates
are from the public releases of C.~Gordon
(\url{https://github.com/chrisgordon1/galactic_bulge_templates}), Pohl
\textit{et al.}~(2022) (\url{https://zenodo.org/records/6276721}), and, for
the Cao \textit{et al.} model, the \texttt{cao13} template distributed with
\texttt{gcepy} (\url{https://github.com/samueldmcdermott/gcepy}).

\newpage 

% Add a Data Availability Statement when the actual access terms are settled.
\bibliography{references}

\clearpage

\appendix
\section{Supplementary Material}

\subsection{Map construction and viewing geometry}

Figure~\ref{fig:maps} shows all six Milky Way analogs. Columns 1--3 follow the DM, old stellar, and quadratic-DM projections of Ref.~\cite{Muru:2025vpz}. Columns 4--5 show the corrected stellar column density $\int\rho_{\star}\,ds$ and the annihilation integral $J=\int\rho_{\rm DM}^{2}\,ds$.

We take the disk normal from the AHF halo-finder profiles used in
Ref.~\cite{Muru:2025vpz}, provided by the authors: the minor axis of the
inertia tensor in the profile bin nearest $10\,h^{-1}$~kpc, which agrees
with the stellar disk normal computed from the particles to within
$2^{\circ}$. The observer is $8$~kpc from the center in the disk plane.
Appendix~A of Ref.~\cite{Muru:2025vpz} places it on the major axis of the
same tensor, but at this radius the profiles give $b/a=0.977$--$0.994$, so
that axis is nearly degenerate: it rotates by $25^{\circ}$ to $82^{\circ}$
between adjacent radial bins of the same profile, and the projected maps,
the stellar map in particular, change substantially over that range. The
prescription therefore does not fix the in-plane azimuth. We choose it for
each galaxy, within this degeneracy, to match the published corner tables;
it lies $10^{\circ}$ to $80^{\circ}$ from the major axis of the
$10\,h^{-1}$~kpc bin. The published axis ratios are then reproduced to
$0.02$ (DM) and $0.05$ (stars), and the published contour levels to within
six percentage points. The remaining residuals reflect the $10^{\circ}$ resolution of our azimuth grid and the choice of $b=0$ crossing described below.
All five panels in a row share this orientation, so their differences from each other arise from the map construction rather than the observer's position.

For the projections of Ref.~\cite{Muru:2025vpz} in columns 1--3, we bin the masses of particles within $15$~kpc of the observer in the flat $(\ell,b)$ coordinates of the original implementation, then convolve with a circular top-hat kernel of radius $3^{\circ}$. 
For the corrected maps in columns 4--5, we replace each particle's mass by $m_i/s_i^{2}$ for stars or $m_i\rho_i/s_i^{2}$ for annihilation before applying the same kernel. 
 This top-hat smoothing is distinct from the Gaussian smoothing used to measure concentration.

The row labels give the bar strength $A_2$, defined as the maximum $m=2$ azimuthal Fourier amplitude of the stellar mass distribution, $|\sum_{i}m_{i}e^{2i\phi_{i}}|/\sum_{i}m_{i}$, where the sum runs over all star particles excluding wind particles, with no age cut. We evaluate this amplitude in $0.5$~kpc cylindrical annuli within $R<6$~kpc and $|z|<3$~kpc of the disk plane.

\subsection{Contour and concentration measurements}

Each panel is normalized to its maximum and shown on a linear color scale, with $\ell$ increasing to the left. Stellar contours are drawn at $70\%$, $50\%$, and $30\%$ of the maximum. The DM contours are matched to the stellar contours where they cross $b=0$. 
A stellar contour crosses $b=0$ at two longitudes, which differ for a
barred map; Ref.~\cite{Muru:2025vpz} does not specify which is used, and
its public code selects, for each level, whichever $b=0$ pixel has the
stellar value nearest that level, on either side of the center. We use
the crossing at positive $\ell$ throughout. The two choices differ by up
to five percentage points in the lowest DM level, and applying the rule
of the public code to our maps changes no axis ratio by more than $0.04$.
The corner tables give axis ratios calculated by singular value decomposition of the enclosed pixel positions, with equal weight per pixel. The axis ratio is the smaller singular value divided by the larger one.

For a nonnegative normalized map, the regions enclosed at fractional levels $q$ and $q^2$ in the linear and squared maps are identical. This explains the repeated axis ratios in columns 1 and 3. The published G1.1 panels of Ref.~\cite{Muru:2025vpz}, for example, give $0.82,0.76,0.68$ at levels $(69\%,52\%,37\%)$ in the linear map and $(47\%,27\%,13\%)$ in the squared map. The displayed percentages are rounded. This identity does not imply equal enclosed flux fractions or concentrations.

For the concentration analysis, we use 36 observer azimuths per halo, separated by $10^{\circ}$, at a galactocentric distance of $8.18$~kpc. The line-of-sight integration extends to $100$~kpc. The resulting 216 views sample six halos, not 216 independent halo realizations. We calculate $C\equiv F(<3^{\circ})/F(<10^{\circ})$ by summing pixels within circular apertures in the flat $(\ell,b)$ plane. With Gaussian full width at half maximum $2^{\circ}$, the annihilation concentrations span $0.342$--$0.496$, with median $0.427$. The stellar concentrations span $0.202$--$0.330$, with median $0.231$. At $3^{\circ}$ Gaussian smoothing, every annihilation map remains more concentrated than the Freudenreich, Cao, and Coleman templates and its corresponding stellar map.

\begin{table}[b]
\caption{Concentration $C\equiv F(<3^{\circ})/F(<10^{\circ})$ at $2^{\circ}$
Gaussian smoothing for the annihilation ($J$) and old-stellar ($\star$) maps
of each halo: the median over its 36 observer azimuths, with the full range
for $J$ in parentheses. The right-hand columns excise the inner
$1.6^{\circ}$, the angular scale of the softening length, from both
numerator and denominator:
$C'\equiv F(1.6^{\circ}<\theta<3^{\circ})/F(1.6^{\circ}<\theta<10^{\circ})$.
The bulge templates and the spherical gNFW$^{2}$ profile are listed for
comparison.}
\label{tab:conc}
\footnotesize
\begin{ruledtabular}
\begin{tabular}{lcccc}
 & $C_{J}$ & $C_{\star}$ & $C'_{J}$ & $C'_{\star}$ \\
\hline
G1.1 & 0.46 (0.44--0.47) & 0.22 & 0.33 (0.32--0.34) & 0.16 \\
G2.1 & 0.35 (0.34--0.36) & 0.21 & 0.25 (0.24--0.26) & 0.15 \\
G3.1 & 0.38 (0.37--0.40) & 0.23 & 0.26 (0.26--0.28) & 0.16 \\
G1.2 & 0.38 (0.36--0.44) & 0.22 & 0.27 (0.25--0.31) & 0.16 \\
G2.2 & 0.46 (0.42--0.50) & 0.27 & 0.33 (0.30--0.35) & 0.19 \\
G3.2 & 0.45 (0.44--0.46) & 0.26 & 0.31 (0.30--0.32) & 0.18 \\
\hline
All 216 views & 0.43 (0.34--0.50) & 0.23 & 0.31 (0.24--0.35) & 0.17 \\
\hline
Freudenreich & 0.27 & & 0.20 & \\
Cao & 0.27 & & 0.19 & \\
Coleman & 0.29 & & 0.21 & \\
gNFW$^{2}$ ($\gamma=1.2$) & 0.51 & & 0.30 & \\
\end{tabular}
\end{ruledtabular}
\end{table}

Table~\ref{tab:conc} lists the concentrations by halo, together with $C'$, the same ratio with the inner $1.6^{\circ}$ excised. The spread among the
36 azimuths of a single halo ($0.01$--$0.08$) is smaller than the spread
among halos (medians $0.35$--$0.46$), so the 216 views are six correlated
samples rather than 216 independent ones. Between $38\%$ and $44\%$ of
$F(<3^{\circ})$ in the annihilation maps arises within $1.6^{\circ}$ of the
center, the angular scale of the softening length, and $33$--$36\%$ for the
stellar maps; these are upper bounds on the unresolved contribution, since
the same lines of sight also traverse well-resolved material at larger
radii. Excising this region from both numerator and denominator lowers
every concentration but changes no ranking: every annihilation map remains
more concentrated than each bulge template and than its own stellar map
(216 of 216) for exclusion radii of $1^{\circ}$, $1.6^{\circ}$, and
$2^{\circ}$, at both $2^{\circ}$ and $3^{\circ}$ smoothing. The conclusion
therefore does not rest on the unresolved center.

The Freudenreich and Coleman templates have not been convolved with the instrumental response. The Cao template is obtained by summing predicted count maps in 14 energy bins, which already include the point-spread function and exposure. Additional Gaussian smoothing therefore does not produce a common effective response. This qualification does not affect the comparisons with the Freudenreich, Coleman, or simulated stellar maps.

For a spherical generalized Navarro--Frenk--White profile with inner slope $\gamma=1.2$ and scale radius $26$~kpc, the gNFW$^{2}$ template, evaluated as the line-of-sight integral of $\rho^{2}$ averaged over each pixel, gives $C=0.51$ at $2^{\circ}$ Gaussian smoothing. 
No view exceeds this concentration; the most concentrated views, six of G2.2, reach $0.49$--$0.50$.

\subsection{Numerical checks}

\begin{figure*}[t]
\includegraphics[width=0.93\textwidth]{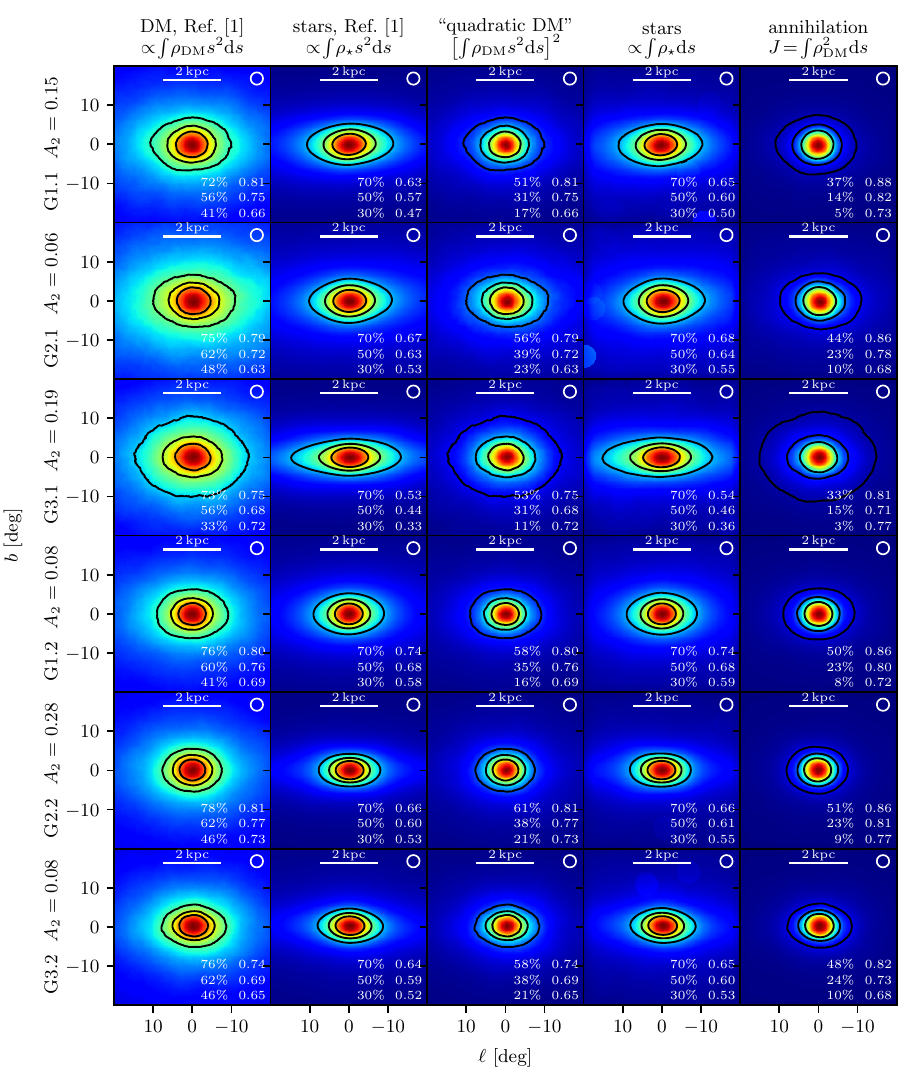}\vskip-0.4cm
\caption{Projected maps of the six Milky Way analogs. Columns 1--3 show the DM mass projection, old stellar mass projection, and squared DM mass projection following Ref.~\cite{Muru:2025vpz}, with the disk normal from the halo-finder profiles of Ref.~\cite{Muru:2025vpz} and the in-plane azimuth chosen, within the  degeneracy of the major axis, to match the published panels of  Ref.~\cite{Muru:2025vpz} (see SM). Columns 4--5 show the corrected stellar column density and annihilation intensity. All panels in a row use the same observer position and include particles out to $15$~kpc from the observer. The corrected maps use the same $3^{\circ}$-radius top-hat aperture as the mass projections. Each panel is normalized to its maximum. The corner tables give the contour levels and axis ratios of the enclosed regions. Stellar levels are $70\%$, $50\%$, and $30\%$, with DM levels matched at $b=0$. Columns 1 and 3 have identical contour regions at fractional levels related by $q\rightarrow q^{2}$. The corrected annihilation maps are more centrally concentrated than the stellar maps. The scale bar is $2$~kpc, and the circle marks the $1.6^{\circ}$ angular scale corresponding to the gravitational softening length.}
\label{fig:maps}
\end{figure*}

We compare direct DM mass sums for G3.1 with the public implementation of Ref.~\cite{Muru:2025vpz}, using the same viewing geometry, $3^{\circ}$ aperture, $15$~kpc maximum distance from the observer, and $401\times401$ grid. 
The maximum relative difference is $5\times10^{-7}$ over all 160,801 lines of sight, with median $2\times10^{-7}$; the residual is the single-precision accumulator of the public code. 
This test covers the linear DM projection; the stellar projection uses the same code path with the old-star particle selection and was not tested separately. 
The binned maps in Fig.~\ref{fig:maps} are computed by histogramming the particles and convolving with the aperture kernel rather than by summing within each aperture directly; 
on G3.1 the two agree to $0.5\%$ (median) and $2\%$ (maximum, at the aperture rim) for the mass projection. 
The disk normal is taken from the halo-finder profiles and the azimuth chosen as described above.

We test the corrected estimator separately by drawing particles from a Navarro--Frenk--White profile and comparing the estimated signal with the analytic $\int\rho^{2}ds$. For G3.1, we vary the number of neighbors over $k=16$--$64$, the maximum line-of-sight distance, and the observer's galactocentric distance. 
We also repeat the calculation using half of the particles. 
We vary the adopted center for G3.2 and the Gaussian smoothing width for all 216 views. 
In each tested case, the annihilation maps remain more concentrated than the bulge templates. 
These checks do not establish numerical convergence with increasing simulation resolution. Smoothing the maps also does not remove the contribution of unresolved central emission to the flux within $3^{\circ}$.

\end{document}